\documentclass{article}
\usepackage[utf8]{inputenc}
\usepackage[margin=1in]{geometry}
\usepackage{setspace}
\usepackage{graphicx}
\usepackage{natbib}
\usepackage{hyperref}
\usepackage{amsmath}
\usepackage{booktabs}
\usepackage{pdfpages}
\hypersetup{colorlinks,allcolors=black}

\title{Large Language Models Can Explain Experts Better Than Experts Themselves}

\author{Mina Cho$^1$, Russell Funk$^1$, Alok Gupta$^1$, Mochen Yang$^1$ \\
$^1$ Carlson School of Management, University of Minnesota\\}
\date{Current Draft: 10/15/2026}

\begin{document}

\doublespacing
\maketitle

\begin{abstract}
Capturing tacit knowledge, or the ``know-how'' embedded in individual experiences and contexts, is a central challenge that limits the preservation and transfer of valuable expertise. Traditional methods rely on experts’ ability to articulate their own reasoning, even though much of tacit knowledge is difficult to verbalize. Here, we show that Large Language Models (LLMs) can uncover tacit knowledge from expert behavior and support knowledge transfer from experts to novices. We first identify prompting strategies that improve LLMs' ability to externalize tacit knowledge (i.e., transform tacit into explicit knowledge), and then evaluate the downstream utility of the resulting LLM-externalized tacit knowledge. Our analysis reveals that LLM-externalized tacit knowledge incorporating expert behavior improves decision quality and helps reduce the expert-novice score gap, outperforming knowledge articulated by human experts. We further show that LLMs meaningfully learn and extract knowledge from expert behavior rather than simply mimicking expert responses, and that these findings generalize across different LLMs and information retrieval methods. These findings reaffirm Polanyi’s Paradox --- that ``\textit{we can know more than we can tell''} -- and suggest that LLMs can help overcome human experts’ articulation bottleneck that has long challenged tacit knowledge externalization. 
\end{abstract}

\section{Introduction}

Tacit knowledge, or the ``know-how” gained through experience, makes up a substantial portion of expertise across a wide range of domains including innovation \citep{tamer_cavusgil_tacit_2003, leonard_role_1998}, manufacturing \citep{kernan_freire_tacit_2023, nakano_engaging_2013}, and drug discovery \citep{ben-menahem_coordinating_2016}. Due to its context-specific and personal nature, tacit knowledge is notoriously difficult to capture \citep{lam_tacit_2000}, limiting the preservation and transfer of valuable expertise. This challenge spans domains and has long been recognized as a central problem in knowledge management \citep{hayes-roth_building_1983, cooke_formal_1986}. In this paper, we demonstrate that Large Language Models (LLMs) can be used to uncover tacit knowledge from expert behavior and support knowledge transfer from experts to novices.

Explicit knowledge can be expressed in words, rules, or other formal representations, whereas tacit knowledge is deeply embedded in individual experiences and contexts and is difficult to fully articulate \citep{smith_role_2001, kothari_use_2012}. As famously stated in Polanyi’s Paradox, {\it “we can know more than we can tell”} \citep{polanyi_tacit_1966}, experts often cannot fully articulate the knowledge underlying their behaviors and decisions \citep{agarwal_knowledge_1990, collins_chapter_2008}. \textit{Tacit knowledge externalization and transfer}, the process of transforming tacit into explicit knowledge and enabling its transfer from experts to novices \citep{benbya_navigating_2024,nonaka_dynamic_1994,nonaka_knowledge-creating_1995}, is therefore a problem at the core of knowledge management. To address this challenge, prior work has developed various methods to capture experts' underlying cognitive processes via different elicitation and formalization techniques, including semi-structured interviews, protocol analysis, concept mapping, and Cognitive Task Analysis (CTA) \citep{clark_cognitive_2008,ryder_integrating_1993,wang_bridging_2024, lebovitz_is_2021, bradley_analyzing_2006, cooke_varieties_1994}. 

While these methods have advanced our understanding of expertise in many domains, they face important limitations. Eliciting knowledge from experts relies on their ability to verbalize their own reasoning, even though much of tacit knowledge is difficult to articulate. In addition, substantial human effort is needed not only to elicit, but also to analyze and organize the collected knowledge into representations suitable for knowledge transfer \citep{hoffman_problem_1987, wright_eliciting_1987}. Consequently, existing tacit knowledge externalization methods face challenges both in the quality of the captured knowledge and their scalability. 

These limitations create widespread costs. Valuable expertise is often not disseminated effectively and may even be lost when experts leave an organization \citep{howells_tacit_1996}. For example, businesses face ``corporate amnesia'', whereby expertise accumulated over years of practice is lost when seasoned employees leave the organization \citep{kransdorf_corporate_1998}. This delays the dissemination of critical expertise, slowing progress in areas where timely knowledge transfer is essential (e.g., drug discovery, scientific innovation, clinical practice \citep{wang_knowledge_2012, lilleoere_knowledge-sharing_2011, greenhalgh_tacit_2008}). Moreover, since tacit knowledge externalization is labor- and time-intensive, organizations can allocate resources to only a small portion of available expertise. Even when knowledge is externalized, the resulting knowledge representations may already be outdated by the time they are documented and shared \citep{schulz_codification_2001}.

Generative AI, and particularly LLMs, offer a unique opportunity to address these challenges  \citep{alavi_knowledge_2024}. Rather than relying on experts to explain their reasoning, LLMs may infer latent decision processes directly from examples of expert behavior. Their potential for knowledge externalization and transfer rests on two key capabilities. First, LLMs can process large amounts of structured and unstructured data to identify latent patterns \citep{benbya_navigating_2024, guo_towards_2024, alavi_knowledge_2024}. Although tacit knowledge is difficult for experts to articulate, it is embedded in their behaviors (e.g., their responses or actions). By learning from these context-dependent behaviors, LLMs can uncover the underlying tacit knowledge that experts use. Second, LLMs can express these latent patterns into concrete decision steps in natural language that can be transferred to novices. LLMs' abilities to learn from experts' decisions and to generate natural language representations of the underlying tacit knowledge distinguish them from other knowledge elicitation methods and position them as a promising tool for tacit knowledge externalization and transfer.

Prior research on LLMs and Knowledge Management has explored how LLMs can support knowledge externalization in different capacities. Some studies use LLMs as knowledge-gathering tools that interact with employees to collect information distributed across the organization \citep{zuin_leveraging_2025,kuks_using_2025}. Other studies leverage LLMs to uncover rules from large-scale training data \citep{wang_why_2025, lu_tacit_2025, schinckus_large_2026}, suggesting that LLMs can identify underlying patterns that can be applied to new contexts. While these studies advance our understanding of LLMs as tools for knowledge externalization, an important question remains: how does LLM-externalized tacit knowledge perform when it comes to supporting downstream decision-making and knowledge transfer to novices?

To address this gap, this study deploys LLMs for knowledge externalization and transfer in a math tutoring context. Specifically, our goal is to externalize the reasoning process (in the form of decision-making steps) that expert math teachers use when responding to student mistakes. Prior methods for capturing tacit knowledge typically involve two stages \citep{fenoglio_tacit_2022, hoffman_eliciting_2006}. The first, knowledge elicitation, involves obtaining knowledge from experts through methods such as semi-structured interviews, think-aloud protocols etc. \citep{fonteyn_description_1993, cooke_varieties_1994}. The second, knowledge representation, formalizes the elicited knowledge into a representation of expert reasoning, such as decision trees, production rules (i.e., if-then rules), or networks of concepts and relationships \citep{wilson_evaluation_2015}. 

Drawing from this two-stage process, we designed prompt manipulations that target analogous components in the LLM-based knowledge externalization process. Manipulation (1), the ``source of tacit knowledge'', targets the first stage by varying whether the LLM is given access to student-tutor conversations (i.e., expert behavior in this context) or must only rely on its training data. Manipulation (2), the ``representation of tacit knowledge'', targets the second stage by varying whether the LLM-generated knowledge is required to have the same number of decision-making steps as expert tutors' own articulated knowledge. To understand how each piece of information affects tacit knowledge externalization, we incrementally apply the manipulations, resulting in three prompt variations (no manipulation, manipulation (1), and manipulation (1)+(2)). Using the externalized tacit knowledge, we conduct two studies to examine their downstream utility: (i) whether it can guide LLMs to generate better tutoring responses and (ii) how it can support knowledge transfer from human experts to human novices (i.e., bridging the expert-novice gap). In both studies, we compare LLM-externalized knowledge to a ``benchmark'' knowledge articulated by human experts. 

We find that LLM-externalized tacit knowledge, when derived from student-tutor conversations, improves both response generation and knowledge transfer. This LLM-externalized tacit knowledge (i) enables LLMs to generate higher-quality tutoring responses than using expert-articulated tacit knowledge and (ii) when presented to novice tutors, reduces the expert-novice score gap relative to expert-articulated guidance. Importantly, these improvements exceed those achieved using expert-articulated tacit knowledge, empirically reaffirming Polanyi’s Paradox \citep{polanyi_tacit_1966}. Mechanism analyses show that LLM-externalized tacit knowledge specifically captures how responses are communicated from experts' observed behaviors. Several robustness checks indicate that LLMs meaningfully learn tacit knowledge from expert conversations rather than just copying their ``style'', and that these findings are consistent across LLM providers and retrieval methods (in-context learning or vector search retrieval). 

We contribute to the literature on knowledge management and organizational learning by proposing and evaluating a scalable approach to tacit knowledge externalization. More broadly, our findings suggest a practical pathway for organizations to leverage expert behavior data to mitigate the articulation bottleneck that limits knowledge externalization and transfer. Our study further identifies important conditions for effective tacit knowledge externalization, including sufficient exposure to behavior data, flexible output structures, and sufficiently capable LLMs. Overall, our study highlights the potential of LLMs as tools for knowledge externalization and transfer, demonstrating downstream benefits for decision-making and novice training.

\vspace{-6pt}

\section{Materials and Methods}

\subsection{Experiment Procedure and Participants}

This study was preregistered (https://osf.io/96nh8). The Institutional Review Board at the University of Minnesota approved the study as exempt research (STUDY00025157). In this study, participants were shown an information page describing the task and provided informed consent before participation. All participants were recruited on Prolific, using prescreening criteria of U.S. residency and an approval rate of over 95\%. Given that we wanted to recruit participants who do not have prior math tutoring experiences (i.e., novice tutors), we excluded teachers through Prolific's employment role filter and also screened out participants who had prior math tutoring experience. During the study, we collected information on non-math tutoring experience and included it as a control variable in our regression analyses. 

Participants were shown short dialogue snippets between a math tutor and a student, each ending with a student mistake. They were instructed to imagine themselves as the tutor and generate a response to the mistake. The study included one practice question followed by five actual questions. Participants were compensated at a rate of \$18.00 per hour. Our experiment consisted of five between-subjects conditions: {\it LLM Baseline, LLM+Both, LLM+Conversation, Human Best Effort, and No Tacit Knowledge}. Participants were not allowed to participate in more than one condition. In the three LLM-externalized knowledge conditions, participants were provided the corresponding externalized tacit knowledge. For the {\it Human Best Effort} condition, participants were shown the expert-articulated knowledge from \cite{wang_bridging_2024}. In the {\it No Tacit Knowledge} condition, participants generated responses without guidance. To help participants understand how to apply the tacit knowledge, participants were shown three example conversations completed with expert teacher responses. For each condition, we identified two steps from each respective tacit knowledge that were common across conditions and provided explanations of how these steps were applied in the example tutor responses (see {\it SI Appendix}, Table S9). The same three example conversations were used across all conditions, while the highlighted steps and explanations varied by condition (see {\it SI Appendix}, Figure S8). In the {\it No Tacit Knowledge} condition, participants were shown the same three examples without explanations.

We conducted a pilot study with 30 participants per condition and later recruited an additional 70 participants per condition, resulting in a total of 100 participants per condition. We excluded participants who failed to follow instructions (e.g., reproduced the tacit knowledge step instead of generating responses). We also excluded participants who failed the attention check question, which asked them to select steps that were not shown as part of the tacit knowledge, indicating insufficient attention paid to the task. The final sample sizes were: {\it LLM Baseline} (85), {\it LLM+Both} (94), {\it LLM+Conversation} (91)\footnote{Three out of five responses given by one participant were excluded because they did not follow the task instructions in these three responses.}, {\it Human Best Effort} (87) and {\it No Tacit Knowledge} (99)

\subsection{Analysis with Alternative Tacit Knowledge}

For (i) domain-agnostic tacit knowledge, we use tips collected from a tutor training webpage titled ``Positive Ways of Correcting Students'' published by Georgia Highlands College,\footnote{https://sites.highlands.edu/tutorial-center/tutor-resources/online-tutor-training/.} offering training modules that are applicable across disciplines. We call this {\it Domain-agnostic knowledge 1}. We also use tips titled ``Classroom Challenge: Handling Wrong Answers'', published by the Center for Research on Learning and Teaching at the University of Michigan,\footnote{https://crlt.substack.com/p/classroom-challenge-handling-wrong} which aims to help instructors improve their teaching practices. We call this {\it Domain-agnostic knowledge 2}. For (ii) out-of-context tacit knowledge, we collect tips on responding to errors in students' writing. We use ``Responding to Errors in Multilingual Students’ Writing: Inclusive Teaching Guides \& Tips'' from Boston University's Teaching Writing website\footnote{https://www.bu.edu/teaching-writing/resources/responding-to-errors-in-multilingual-students-writing/} ({\it Out-of-context knowledge 1}) and ``Tips for Writing Tutors: Ten Tips for English as a Second Language Tutorials'' from The University of North Carolina at Chapel Hill's Writing Center\footnote{https://writingcenter.unc.edu/tips-for-writing-tutors/} ({\it Out-of-context knowledge 2}). For (iii) domain-adapted tacit knowledge, we mechanically replace writing-related terms in {\it Out-of-context knowledge 1} and {\it Out-of-context knowledge 2} with math-related terms (e.g., ``writing'' to ``problem-solving'' and ``language problem'' to ``mathematical problem''). We call these {\it Domain-adapted knowledge 1} and {\it Domain-adapted knowledge 2}.

\vspace{-6pt}

\subsection{Validation of Evaluation Criteria Using Experienced Human Tutors}

We constructed two sets of unique conversations (Set A and B) for evaluation by experienced human tutors. Each conversation contained a dialogue between a math tutor and a student, the student's mistake, and the tutor's response to the mistake. Each set contained 20 conversations, all from the 122 complete testing conversations. Two undergraduate tutors evaluated Set A, and two different undergraduate tutors evaluated Set B. Four PhD-student tutors evaluated all 40 conversations across both sets. Each set included a mix of conversations with responses from the LLM-externalized knowledge conditions, {\it No Tacit Knowledge}, {\it Human Best Effort}, novices, and experts, with at least one response from each condition. These responses were from the main analysis, generated using GPT-4o (excluding responses from novices and experts). To ensure that the two sets contained conversations with a similar range of response quality, we allocated conversations such that the distributions of usefulness, caring, and human-sounding scores, evaluated by GPT-4o, had similar standard deviations across Sets A and B. Tutors were blinded to the condition associated with each response. Before conducting the evaluation tasks, tutors were provided with the definitions of usefulness, caring, and human-sounding as reported in {\it SI Appendix}, Table S2. Tutors were then asked to rate each response from 0--10 on each of the three criteria. Undergraduate tutors were compensated $\$30$ and PhD tutors $\$60$ for their participation.

To ensure inter-rater reliability, we calculated intraclass correlation coefficients (ICCs) separately for each set \cite{koo_guideline_2016}.  For Set A, the ICC for overall response quality was ICC(2,6) = 0.81, with criterion-specific ICCs of usefulness (0.82), caring (0.81), and human-sounding (0.64). For Set B, the ICC for overall response quality was ICC(2,6) = 0.77 with criterion-specific ICCs of usefulness (0.82), caring (0.78), and human-sounding (0.58). This shows that the reliability was good for overall response quality in both sets, with criterion-specific reliability ranging from moderate to good. To assess Spearman and Pearson correlations, ratings from the six tutors assigned to each response were averaged within each evaluation criterion to obtain a mean human rating for usefulness, caring, and human-sounding for each conversation. We then compared these human ratings with the corresponding GPT-4o ratings, averaged across the ten iterations.

\begin{figure*}[t!]
\centering
\includegraphics[scale=0.65]{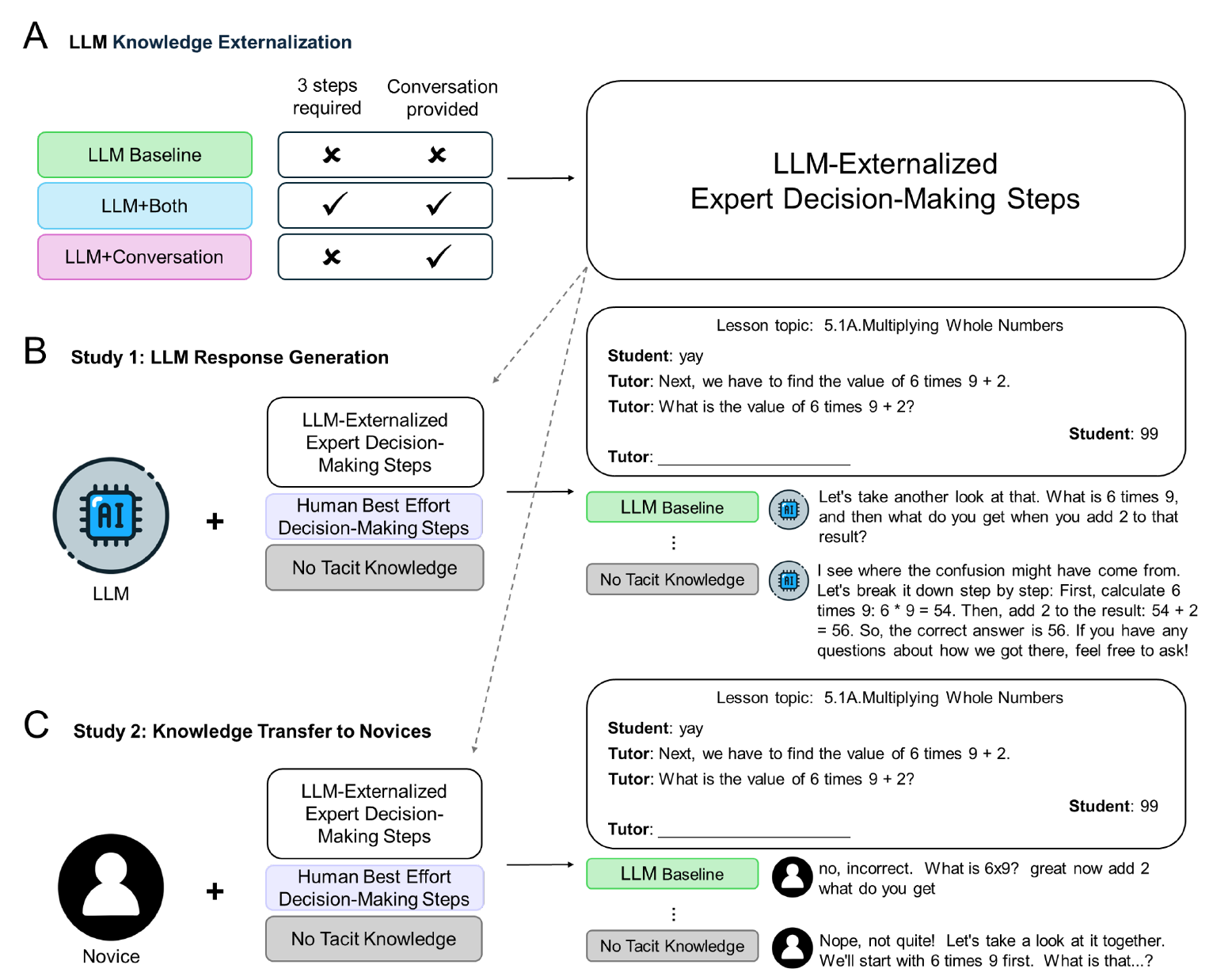}
\caption{Overview of the study design. Panel A shows how a large language model (LLM), GPT-4o, externalizes the tacit knowledge: \textit{LLM Baseline}, externalized by the LLM without example tutoring conversations; \textit{LLM+Conversation}, externalized from tutoring conversations containing novice and expert responses, without a three-step requirement; and \textit{LLM+Both}, externalized from such conversations with a three-step requirement. Dashed arrows indicate that these tacit knowledge representations were subsequently used in Studies 1 and 2. In both studies, \textit{Human Best Effort} supplies the three-step framework developed through interviews with expert tutors in prior research, and \textit{No Tacit Knowledge} supplies no added guidance. Panel B depicts Study 1, in which GPT-4o generates responses to students' mistakes using the assigned guidance. Panel C depicts Study 2, in which human novice tutors receive the assigned guidance during training before writing responses to students' mistakes. Panels B and C each show an example conversation and responses from two conditions.}
\label{fig:flowchart}
\end{figure*}

\section{Results}

\subsection{Data} 
Our objective is to understand how LLMs can be used to externalize experts’ tacit knowledge and facilitate its transfer to novices. We use the dataset collected by \cite{wang_bridging_2024}, consisting of real-world math tutoring conversations between first- to fifth-grade students and tutors. Each conversation includes the final four dialogue turns ending with a student's mistake, followed by a tutor's response. In the original dataset, these responses are provided by novice tutors. The authors of \cite{wang_bridging_2024} then asked expert math tutors how they would have addressed the same mistake, effectively offering an expert-level alternative response for each mistake. The dataset includes 700 conversations, which \cite{wang_bridging_2024} divide into 420 conversations for training, 70 for validation, and 210 for testing (for building their decision-making framework). An example conversation is provided in {\it SI Appendix}, Figure S1. 

Importantly, to capture how expert teachers respond to students' mistakes, \cite{wang_bridging_2024} conducted CTA through interviews with four expert math tutors and developed a three-step decision-making model. This model is a representation of experts' tutoring knowledge, and we refer to it as the “human best effort benchmark”, as it reflects a framework developed by experts articulating their own tacit knowledge. The “human best effort benchmark” is reproduced in {\it SI Appendix}, Figure S2. 

To identify prompting strategies that improve LLMs’ ability to externalize tacit knowledge, we design three prompting strategies that ask an LLM to ``extract the thinking process of expert math teachers'', but vary the information available to the LLM during the externalization process. Specifically, we manipulate (1), the ``source of tacit knowledge'', varying whether the LLM is given access to student-tutor conversations and (2), the ``representation of tacit knowledge'', varying whether the LLM-generated knowledge is required to have the same number of decision-making steps as the knowledge representation in the ``human best effort benchmark''. We incrementally apply the two manipulations, resulting in three conditions: {\it LLM Baseline} (no student-tutor conversations provided), {\it LLM+Conversation} (student-tutor conversations provided without knowledge structure requirement), and {\it LLM+Both} (student-tutor conversations provided and 3-step structure required).

We used OpenAI's GPT-4o\footnote{Given its knowledge cutoff of September 30, 2023, GPT-4o could not have accessed \cite{wang_bridging_2024}, which was first released on arXiv on October 16, 2023, thereby mitigating concerns about data leakage.} for tacit knowledge externalization in the main analysis, and later replicated the analysis with other LLMs (see ``Replication with Other LLMs''). Figure \ref{fig:flowchart}'s Panel A summarizes the experimental conditions, and the prompts are reported in {\it SI Appendix}, Table S1. 

In the {\it LLM+Conversation} and {\it LLM+Both} conditions, the LLM is given access to student-tutor conversations during knowledge externalization\footnote{Conversations include the following: conversation ID, lesson topic, conversation history ending with the student's mistake, novice tutor's response, and expert teacher's response.}. We randomly sample 320 conversations from the full training set of 420, as this is the maximum number of conversations that could fit into GPT-4o's context window.\footnote{In Section ``Replication with Retrieval-Augmented Generation (RAG)'', we consider an alternative setting where all 420 training conversations are made available to the LLM via Retrieval-Augmented Generation (RAG). Key findings are consistent across the two settings.} The LLM's generation temperature is set to 0 to minimize stochastic variation. Nevertheless, identical prompts can still produce slightly different knowledge representations due to the LLM's inherent stochasticity \cite{rathje_gpt_2024}. We therefore repeat the externalization process ten times for each condition. Examples of externalized tacit knowledge for each condition are shown in {\it SI Appendix}, Figures S3-S5.

\subsection{Study 1. Response Generation}

Study 1's objective is to examine whether the externalized tacit knowledge can guide an LLM to make more expert-like decisions. To do so, we prompt the LLM to ``imagine you are an expert math teacher'' and generate responses to student mistakes for testing conversations (unseen during knowledge externalization). As part of the prompt for response generation, we provide the LLM with different variations of externalized tacit knowledge, describing it as ``tacit knowledge that expert math teachers use to determine initial responses to students' mistakes''. We repeat the response generation process ten times for each condition, using each of the externalized tacit knowledge. 

Out of the 280 testing conversations (we combine \cite{wang_bridging_2024}'s validation and testing conversations to construct our testing set), 122 are ``complete'' in the sense that the original math problems are clearly articulated within the four messages leading up to the student’s mistake. For other conversations, there is insufficient information to infer the original problems. We only use these ``complete” conversations for response evaluation to ensure that the LLM has enough context to generate reasonable responses.

In addition to the three LLM-externalized tacit knowledge conditions, we added two conditions for comparison purposes. First, {\it No Tacit Knowledge} serves as a baseline in which the LLM generates responses without any tacit knowledge (i.e., a ``zero-shot'' setting). Second, {\it Human Best Effort} provides the LLM with the ``human best effort benchmark'' articulated by experts. Figure \ref{fig:flowchart}'s Panel B shows two examples of LLM-generated responses. 

We evaluate the quality of LLM-generated responses using three criteria adapted from \cite{wang_bridging_2024}: (i) usefulness,\footnote{The original definition of usefulness in \cite{wang_bridging_2024} did not explicitly penalize responses that directly provided correct answers to students. To better capture responses that are useful for student \textit{learning}, we amended the usefulness definition with the following sentence: ``Useful responses are responses that encourage students to recognize the mistake on their own and guide students to self-correct, rather than directly providing the answer.''} (ii) caring, and (iii) human-sounding. Definitions of all three criteria are provided in the {\it SI Appendix}, Table S2. Each response is rated on a 0--10 scale per criterion using GPT-4o (and later replicated with different scoring LLMs in ``Replication with Other LLMs''), and we compute an overall score as the average across the three criteria. We use chain-of-thought prompting, instructing the LLM to articulate its reasoning before scoring \cite{scarlatos_training_2025}. The prompt used for scoring is reported in the {\it SI Appendix}, Table S1. For each condition, response quality is averaged across the ten repetitions. 

\subsubsection{LLM-Externalized Tacit Knowledge Generates Effective Responses}

\begin{figure*}[!t]
\centering
\includegraphics[scale=0.3]{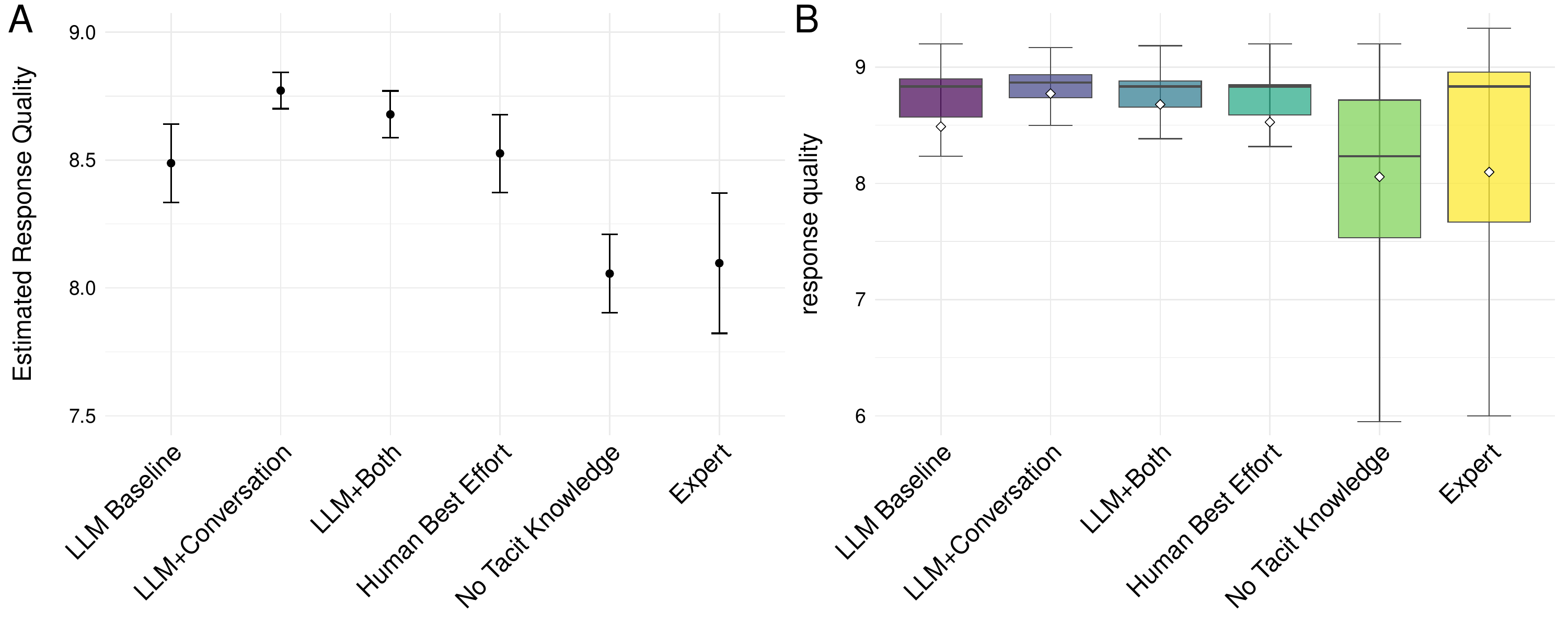}
\caption{Quality of responses under different tacit knowledge conditions. All conditions except \textit{Expert} contain responses to students' mistakes generated by the large language model (LLM), GPT-4o. Condition names identify the tacit knowledge supplied: \textit{No Tacit Knowledge}, no added guidance; \textit{Human Best Effort}, the three-step framework developed through interviews with expert tutors in prior research; \textit{LLM Baseline}, externalized by the LLM without example tutoring conversations; \textit{LLM+Conversation}, externalized from tutoring conversations containing novice and expert responses, without a three-step requirement; and \textit{LLM+Both}, externalized from such conversations with a three-step requirement. \textit{Expert} denotes the original expert-written responses. Overall response quality is the average of usefulness, caring, and human-sounding ratings. Panel A shows estimated marginal means and 95\% CIs from a LMM accounting for conversation-level clustering and condition-specific residual variances. Panel B shows the distribution of response quality; diamonds mark condition means.}
\label{fig:response_generation}
\end{figure*}

Figure \ref{fig:response_generation}'s Panel A reports the response quality of LLM-generated responses across the five conditions. We also report the response quality of expert human tutors (labeled {\it Expert}) evaluated on the same set of ``complete'' testing conversations. 

We first tested whether the presence of tacit knowledge affects response quality. We used a linear mixed-effects model (LMM) to predict response quality from the fixed effect of condition ({\it No Tacit Knowledge, LLM Baseline, LLM+Conversation, LLM+Both, Human Best Effort, Experts}), with conversation ID included as a random intercept to account for the non-independence of responses generated for the same conversation. Because residual variability differed across conditions, we allowed condition-specific residual variances. There was a statistically significant difference of response quality between conditions as determined by one-way analysis of variance (ANOVA) (\(F\)(5, 605) = 35.01, \(p < 0.0001\)). A Tukey’s honestly significant difference (HSD) test revealed that compared to responses from {\it No Tacit Knowledge}, responses were higher-quality when tacit knowledge was given. This was true regardless of whether tacit knowledge was externalized by LLMs or articulated by human experts; {\it Human Best Effort} ($\Delta M$ = 0.470, \(p < 0.0001\)), {\it LLM Baseline} ($\Delta M$ = 0.432, \(p = 0.0001\)), {\it LLM+Conversation} ($\Delta M$ = 0.716, \(p < 0.0001\)), and {\it LLM+Both} ($\Delta M$ = 0.623, \(p < 0.0001\)).

Moreover, responses generated from two out of the three LLM-externalized tacit knowledge conditions significantly outperformed those from {\it Expert}. Tukey's HSD test showed that responses from {\it LLM+Conversation} ($\Delta M$ = 0.675, \(p < 0.0001\)), and {\it LLM+Both} ($\Delta M$ = 0.582, \(p = 0.0004\)) performed better 
compared to human experts' responses. Figure \ref{fig:response_generation}'s Panel B also shows that LLM-externalized tacit knowledge conditions result in more \textit{stable} response quality across conversations (indicated by the narrower distributions) compared to {\it Expert}. 

LLM-externalized tacit knowledge can enable LLMs to generate responses that not only match but also exceed the quality of responses based on expert-articulated tacit knowledge. Tukey's HSD tests showed {\it LLM Baseline} (\(p = 0.999\)) and {\it LLM+Both} (\(p = 0.312\)) were not significantly different from {\it Human Best Effort}, and {\it LLM+Conversation} significantly outperforms it ($\Delta M$ = 0.246, \(p = 0.0044\)). In other words, incorporating expert conversations and allowing a flexible output structure during LLM externalization generates higher-quality responses than prompting the LLM with expert-articulated tacit knowledge. These findings are consistent with Polanyi's Paradox: while experts possess rich tacit knowledge, they often struggle to articulate it effectively \cite{polanyi_tacit_1966}.

\begin{figure*}[t!]
\centering
\includegraphics[scale=0.45]{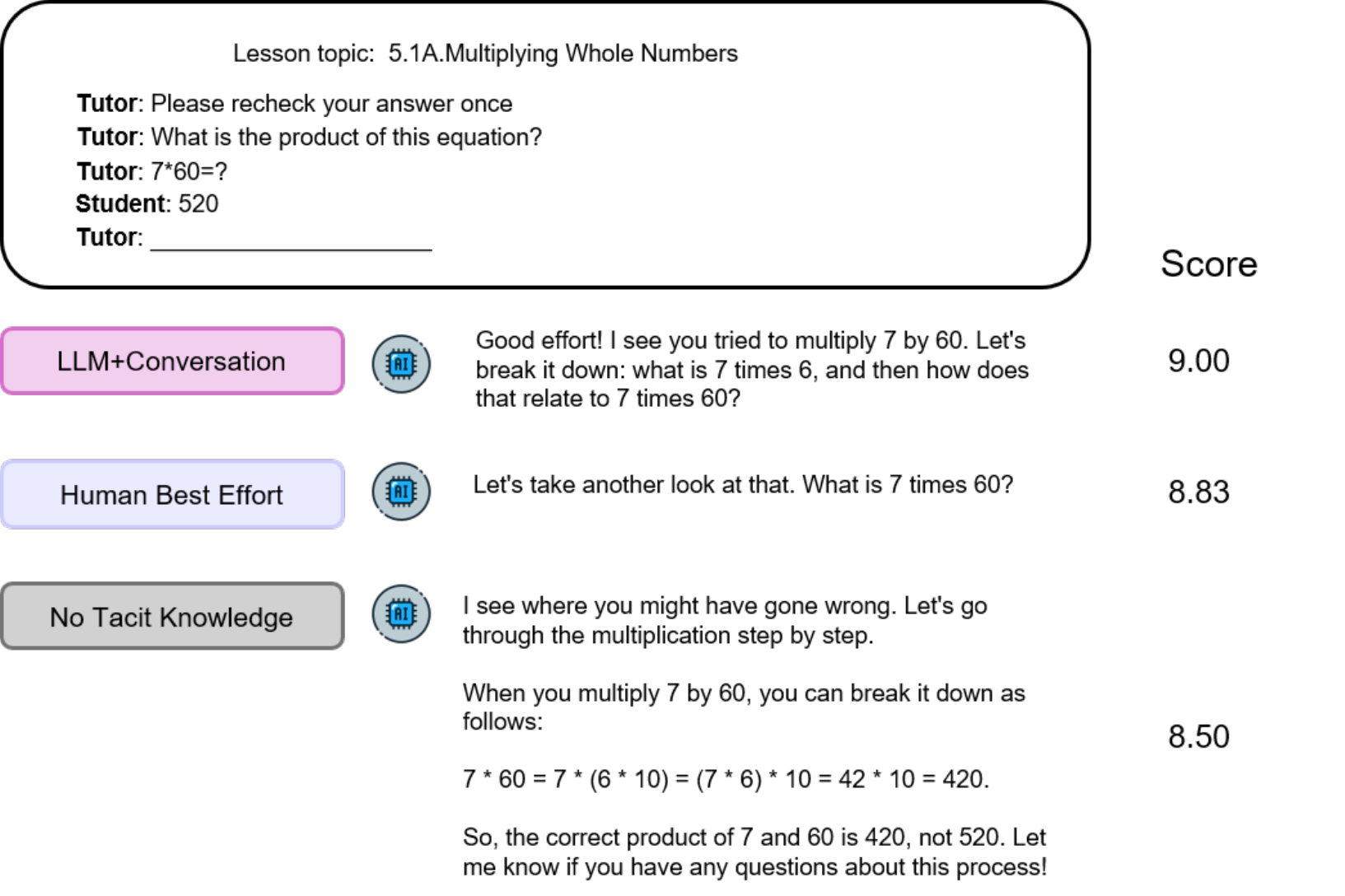}
\caption{Example tutor responses to a student error and their corresponding overall scores. All three responses were generated by GPT-4o, a large language model (LLM). \textit{LLM+Conversation} supplies tacit knowledge externalized from tutoring conversations containing novice and expert responses, without a three-step requirement; \textit{Human Best Effort} supplies the three-step framework developed through interviews with expert tutors in prior research; and \textit{No Tacit Knowledge} supplies no added guidance. Overall response quality is the average of usefulness, caring, and human-sounding ratings. The example illustrates what the observed score differences translate to in practice.}
\label{fig:example_response}
\end{figure*}

 To illustrate what the observed score differences translate to in practice, Figure \ref{fig:example_response} presents an example student-tutor conversation ending in a student error, along with LLM-generated responses and their corresponding overall scores. We present three conditions: {\it LLM+Conversation}, which produced the highest-quality responses on average; {\it Human Best Effort}, which uses knowledge articulated by human experts; and {\it No Tacit Knowledge}, which serves as a baseline. The response generated under {\it LLM+Conversation} provides supportive encouragement (``Good effort!'') and guides the student to fix the error in a way that helps them solve the problem correctly in the future. The response generated by {\it Human Best Effort} restates the problem without offering meaningful guidance, while {\it No Tacit Knowledge} directly gives away the answer. Additional example conversations and responses are shown in {\it SI Appendix}, Figures S6--S7.  

Next, we look at how each incremental change in knowledge externalization prompting design affects response quality. Providing expert conversations during externalization substantially improves downstream performance, demonstrated by how {\it LLM+Conversation} outperforms {\it LLM Baseline} ($\Delta M$ = 0.284, \(p < 0.001\)). By contrast, additionally imposing restrictions on knowledge structure \textit{reduces} response quality; {\it LLM+Both} performs worse than {\it LLM+Conversation} ($\Delta M$ = 0.093, \(p < 0.05\)). 

As a robustness check, we repeated our analysis using pairwise paired-sample \(t\)-tests with adjustments for multiple comparisons (see {\it SI Appendix}, Table S3). The results are consistent with those obtained from the mixed-effects model.

These results suggest that providing experts' actual behaviors is critical for LLMs to effectively externalize tacit knowledge (which leads to high-quality downstream decisions). However, imposing a predefined output structure that mimics expert-articulated knowledge turns out to be counterproductive. Allowing greater flexibility is important for LLMs to recognize patterns and capture the nuances of expert tacit knowledge from expert behaviors.

\vspace{-6pt}

\subsection{Study 2. Knowledge Transfer}
Next, we study how LLM-externalized tacit knowledge can support knowledge transfer from experts to novices. We conduct an experiment on Prolific where novice tutors generate responses to students’ mistakes. Participants are randomly assigned to conditions that differ in the type of tacit knowledge provided to guide their response generation \footnote{
Although the externalized knowledge was generally consistent across the ten iterations, some variations in wording, step order, and the number of steps (for {\it LLM+Conversation} and {\it LLM Baseline}) were observed. For each condition, we selected a single representative version of tacit knowledge from the ten iterations to use in the experiment. They are shown in {\it SI Appendix}, Figures S3-S5}.

A screenshot of the experiment is presented in {\it SI Appendix}, Figure S8. After reviewing the tacit knowledge guidelines, each participant provided responses to five conversations ending in a student's mistake, randomly sampled from 104 conversations (from 122 ``complete” testing conversations after removing 18 duplicates)\footnote{Duplicates resulted from \cite{wang_bridging_2024} asking more than one expert teacher to respond to the same error.}. Figure \ref{fig:flowchart}'s Panel C shows example responses generated by novice tutors for two conditions. Additional details of the experiment are reported in the Materials and Methods section. Descriptive statistics for participants are presented in {\it SI Appendix}, Table S4. Since all responses from Study 2 are generated by human novice tutors, we evaluate the quality of responses using (i) usefulness and (ii) caring, and exclude (iii) human-sounding.\footnote{Results are qualitatively similar when using three evaluation criteria (See {\it SI Appendix}, Figure S9).}

\subsubsection{LLM-Externalized Tacit Knowledge Bridges the Expert-Novice Gap}

\begin{figure}
\centering
\includegraphics[width=.5\linewidth]{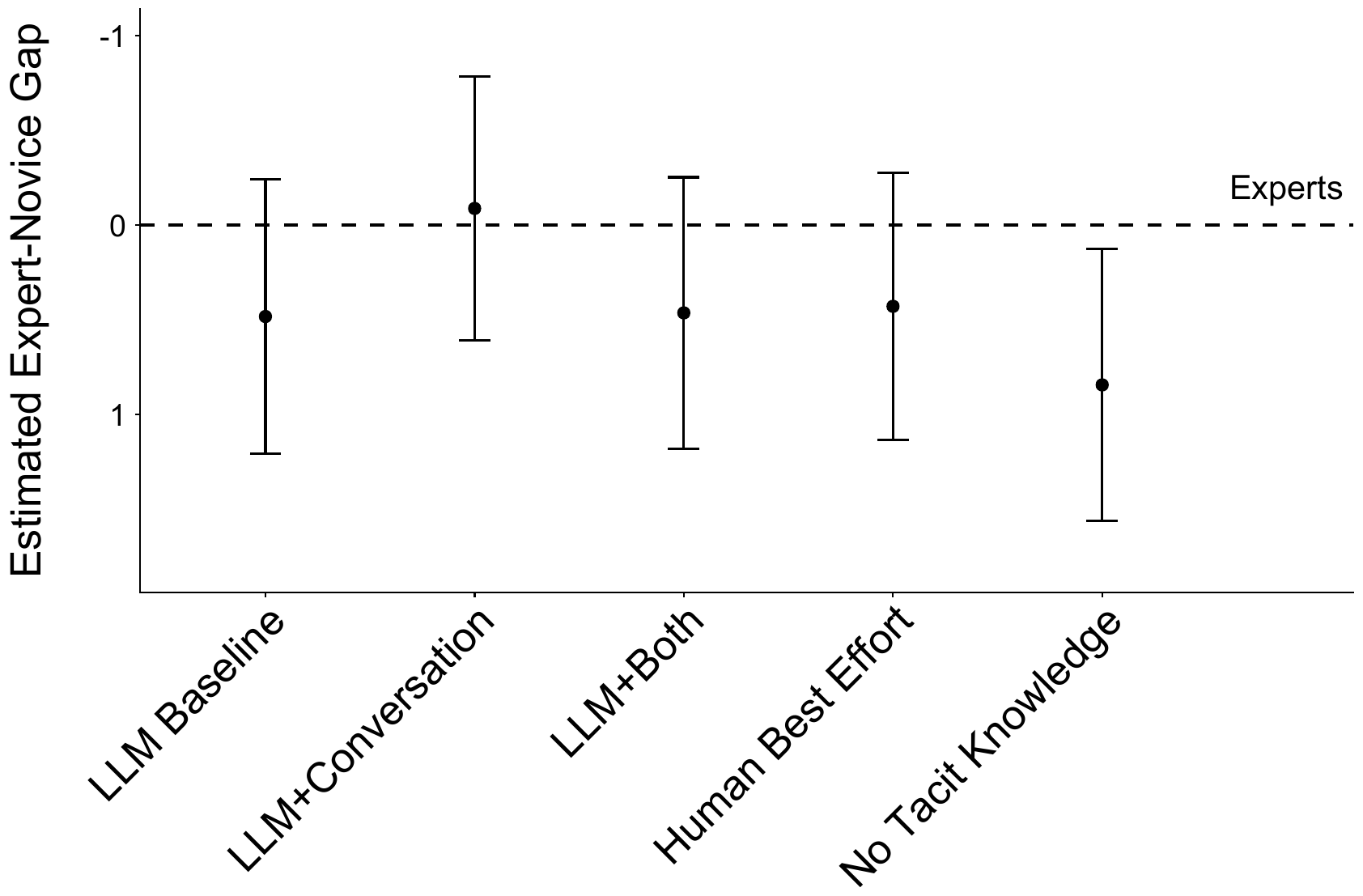}
\caption{Expert--novice gap after knowledge transfer. Human novices in all five conditions wrote responses to students' mistakes. LLM denotes large language model. Condition names identify the tacit knowledge supplied during training: \textit{No Tacit Knowledge}, no added guidance; \textit{Human Best Effort}, the three-step framework developed through interviews with expert tutors in prior research; \textit{LLM Baseline}, externalized by the LLM without example tutoring conversations; \textit{LLM+Conversation}, externalized from tutoring conversations containing novice and expert responses, without a three-step requirement; and \textit{LLM+Both}, externalized from such conversations with a three-step requirement. The expert--novice gap is the expert response score minus the novice response score for the same conversation, averaged across usefulness and caring. Positive values favor experts; negative values favor novices. Points show estimated marginal means, and error bars show 95\% CIs from a LMM accounting for conversation- and participant-level clustering and condition-specific residual variances. Each participant responded to five conversations. The dashed line labeled \textit{Experts} marks a gap of zero; points closer to this line indicate smaller differences in average response quality between experts and novices.}
\label{fig:knowledge_transfer_inverted}
\end{figure}

Figure \ref{fig:knowledge_transfer_inverted} presents the results of the online experiment. For each conversation, we have two responses, respectively generated by a tacit knowledge-trained novice and by a human expert. We define the {\it expert-novice gap} as the difference in response quality between these two responses,\footnote{For the 18 duplicate conversations, we retain the conversation with the higher expert response quality, providing a conservative estimate of the reduction in the expert-novice gap.} and it is calculated as the average of the differences in usefulness and caring scores. The dashed line labeled ``Expert'' corresponds to a value of zero (indicating no gap). Thus, points closer to the dashed line indicate smaller expert-novice performance gaps.

We used an LMM to predict the expert-novice gap from the fixed effect of condition ({\it No Tacit Knowledge, LLM Baseline, LLM+Conversation, LLM+Both, Human Best Effort}). We include conversation ID and Participant ID (we give unique Participant ID across all conditions) as random intercepts to account for the non-independence of responses generated from the same conversation and participant. We include covariates of question order (to account for potential fatigue or primacy effects), non-math tutoring experience, sex, age, and education. Because residual variability differed across conditions, we allowed condition-specific residual variances.

There was a statistically significant difference of expert-novice gap between conditions as determined by one-way ANOVA (\(F\)(4, 2154) = 12.57, \(p < 0.0001\)). {\it No Tacit Knowledge} has a significantly positive expert-novice gap ($M$ = 0.843, 95\% confidence interval (CI) [0.126, 1.561]), suggesting that with no training, novices produce lower-quality responses. However, when novices are trained with externalized knowledge, it results in a narrower expert-novice gap. The estimated expert–novice gap does not significantly differ from zero for conditions where tacit knowledge was given\footnote{Although these estimated gaps are not statistically distinguishable from zero, the confidence intervals do not establish equivalence between novice and expert performance}: {\it LLM Baseline} ($M$ = 0.482, 95\% CI [-0.242, 1.207]), {\it LLM+Conversation} ($M$ = -0.088, 95\% CI [-0.784, 0.607]), {\it  LLM+Both} ($M$ = 0.463, 95\% CI [-0.253, 1.180]), and {\it Human Best Effort} ($M$ = 0.429, 95\% CI [-0.277, 1.134]).

Tukey's HSD tests show that {\it LLM+Conversation} results in a significantly smaller expert-novice gap compared to all other conditions; {\it LLM Baseline} ($\Delta M$ = 0.571, \(p <0.01\)), {\it LLM+Both} ($\Delta M$ = 0.552, \(p < 0.001\)), {\it Human Best Effort} ($\Delta M$ = 0.517, \(p <0.01\), and {\it No Tacit Knowledge} ($\Delta M$ = 0.932, \(p <0.0001\)). Interestingly, among the three LLM-externalized tacit knowledge conditions, only {\it LLM+Conversation} creates a significantly smaller expert-novice gap than {\it Human Best Effort}. Tukey's HSD comparisons show that there is no statistically significant difference between {\it Human Best Effort} and {\it LLM Baseline} (\(p = 0.997\)) or {\it LLM+Both} (\(p = 0.999\)). Tacit knowledge from these two LLM-based conditions does no better at closing the expert-novice gap than human-articulated tacit knowledge.

As a robustness check, we repeated our analysis using a multiple regression model, with the Expert–Novice Gap as the dependent variable and experimental condition as the independent variable (see {\it SI Appendix}, Tables S5 and S6). The results are substantively consistent with those obtained from the mixed-effects model.

Overall, {\it LLM+Conversation} produced the strongest knowledge-transfer outcome among the externalized tacit knowledge. Compared with {\it Human Best Effort}, it reduced the expert-novice score gap by an estimated 0.517 points (Tukey-adjusted \(p <0.01\)). This improvement during a brief intervention highlights the potential of LLM-derived guidance to improve novices' tutoring responses.

\section{Mechanism Analysis}

Across Studies 1 and 2, tacit knowledge externalized under the {\it LLM+Conversation} condition is the most effective in both guiding LLMs to generate high-quality responses and  facilitating knowledge transfer to novices. To understand what LLMs can externalize that human experts struggle to articulate, we compare the {\it LLM+Conversation} and {\it Human Best Effort} conditions.  To unpack the differences between the two conditions, we decompose overall response quality into its constituent components.

\subsubsection{Study 1 Decomposed: LLM+Conversation Improves Caring and Human-Sounding Scores}
 
{\it SI Appendix}, Figure S10 presents a breakdown of response quality scores from Study 1. {\it LLM+Conversation} achieves significantly higher scores than {\it Human Best Effort} in 
caring ($\Delta M$ = 0.448, \(p < 0.0001\)) and human-sounding ($\Delta M$ = 0.119, \(p < 0.01\)), with a marginally significant improvement in usefulness ($\Delta M$ = 0.170, \(p = 0.063\)). These findings suggest that, compared to expert-articulated knowledge, providing LLMs with experts' observed behaviors during knowledge externalization generates responses that are more caring and human-sounding, with a trend toward greater usefulness. These results suggest that the advantage of {\it LLM+Conversation} over {\it Human Best Effort} is clearest in how responses are communicated. 

\subsubsection{Study 2 Decomposed: LLM+Conversation Guides Novices to Write More Caring Responses}

{\it SI Appendix}, Figure S11 presents a breakdown of the expert-novice gap across conditions. {\it LLM+Conversation} is the most effective at closing the gap for both components in absolute terms. For usefulness, Tukey's HSD tests show that the difference between {\it LLM+Conversation} and {\it Human Best Effort} is not statistically significant ($p = 0.646$). In contrast, in terms of caring, novices in the {\it LLM+Conversation} condition achieve a significantly smaller expert-novice gap than those in the {\it Human Best Effort} condition ($\Delta M$ = 0.818, \(p < 0.001\)). Compared to expert-articulated knowledge, providing the LLM with experts' observed behaviors results in tacit knowledge that better trains novices in generating caring responses.

To better understand this result, we conduct an exploratory analysis of the externalized tacit knowledge. In {\it LLM+Conversation}, the ``caring'' component is explicitly highlighted in two different decision steps of the tacit knowledge representation (see {\it SI Appendix}, Figure S4). Step 2, ``Acknowledge the Student's Effort'', instructs the tutor to begin the response by acknowledging the student's effort and attempt, even if it is incorrect. Step 8, ``Provide Positive Reinforcement'', emphasizes ending the interaction with encouragement. These steps make ``caring'' an explicit consideration in response generation. 

In comparison, expert-articulated tacit knowledge does not have the ``caring'' component as a distinct decision step ({\it SI Appendix}, Figure S2). Instead, it appears as one of several possible options under a high-level decision step. Specifically, Step 2, ``Determine the Strategy'', and Step 3, ``Identify the Intention behind the strategy'', each encompasses ten possible options, and ``Encourage the student'' and ``Motivate the student'' are just two of those options. As a result, the caring aspect is much less evident in expert-articulated tacit knowledge. A similar pattern appears in the tacit knowledge externalized in {\it LLM+Both}, where no step explicitly targets the ``caring'' component. 

Interestingly, like {\it LLM+Conversation}, externalized tacit knowledge in {\it LLM Baseline} also has two steps that explicitly relate to the ``caring'' component ({\it SI Appendix}, Figure S3). In particular, Step 7, ``Consider the Student's Confidence and Motivation'', instructs the tutor to think about how the response might affect the student's confidence and motivation. Step 8, ``Plan the Tone and Approach'', instructs the tutor to choose a tone that is supportive and encouraging. However, responses in {\it LLM Baseline} have a similar gap on the caring component as {\it Human Best Effort} and {\it LLM+Both} conditions.

We hypothesize that the discrepancy may be a result of the positions and the concreteness of the caring-related steps. Although both conditions have two steps that explicitly instruct tutors to be caring, the steps are featured in more prominent positions in {\it LLM+Conversation} (Steps 2 and 8 out of eight steps) compared to {\it LLM Baseline} (Steps 7 and 8 out of nine steps). Also, Step 2 in {\it LLM+Conversation} is highly concrete and ``executable'' -- beginning the response with a comment that acknowledges the student's effort is straightforward for novices to follow. As illustrative evidence, we manually inspect the responses from the two conditions and find that around 60\% of responses in the {\it LLM+Conversation} condition begin with a caring statement such as ``Great job'' or ``Good try'', compared to around 40\% of responses in {\it LLM Baseline}.

\section{Additional Analyses and Robustness Checks}

Here, we conduct several additional analyses to verify the robustness of the main results from Study 1 and Study 2.

\subsection{Varying the Number of Conversations used for Knowledge Externalization}
Results in Study 1 showed that {\it LLM+Conversation} yields the highest-quality LLM-generated responses, demonstrating that expert conversations are necessary for effective knowledge externalization. If LLMs meaningfully {\it learn} and {\it extract} knowledge from expert conversations (rather than simply copying their ``style''), the quality of externalized knowledge, and in turn, generated responses, should depend on whether a sufficient number of expert conversations are given. To verify that LLMs meaningfully {\it learn} and {\it extract} knowledge from expert conversations, we vary the number of conversations available during knowledge externalization and assess the quality of the generated responses \cite{agarwal_many-shot_2024}. 

We begin with the entire 320 training conversations previously used for knowledge externalization and construct smaller samples through nested subsampling: 300 conversations are randomly sampled from the 320-conversation set, 200 from the 300, 100 from the 200, 50 from the 100, and so on down to a single conversation. This nested design ensures that increases in sample size correspond to the addition of new conversations, showing us how response quality changes as more conversations become available for knowledge externalization.

\begin{figure}
\centering
\includegraphics[width=0.55\linewidth]{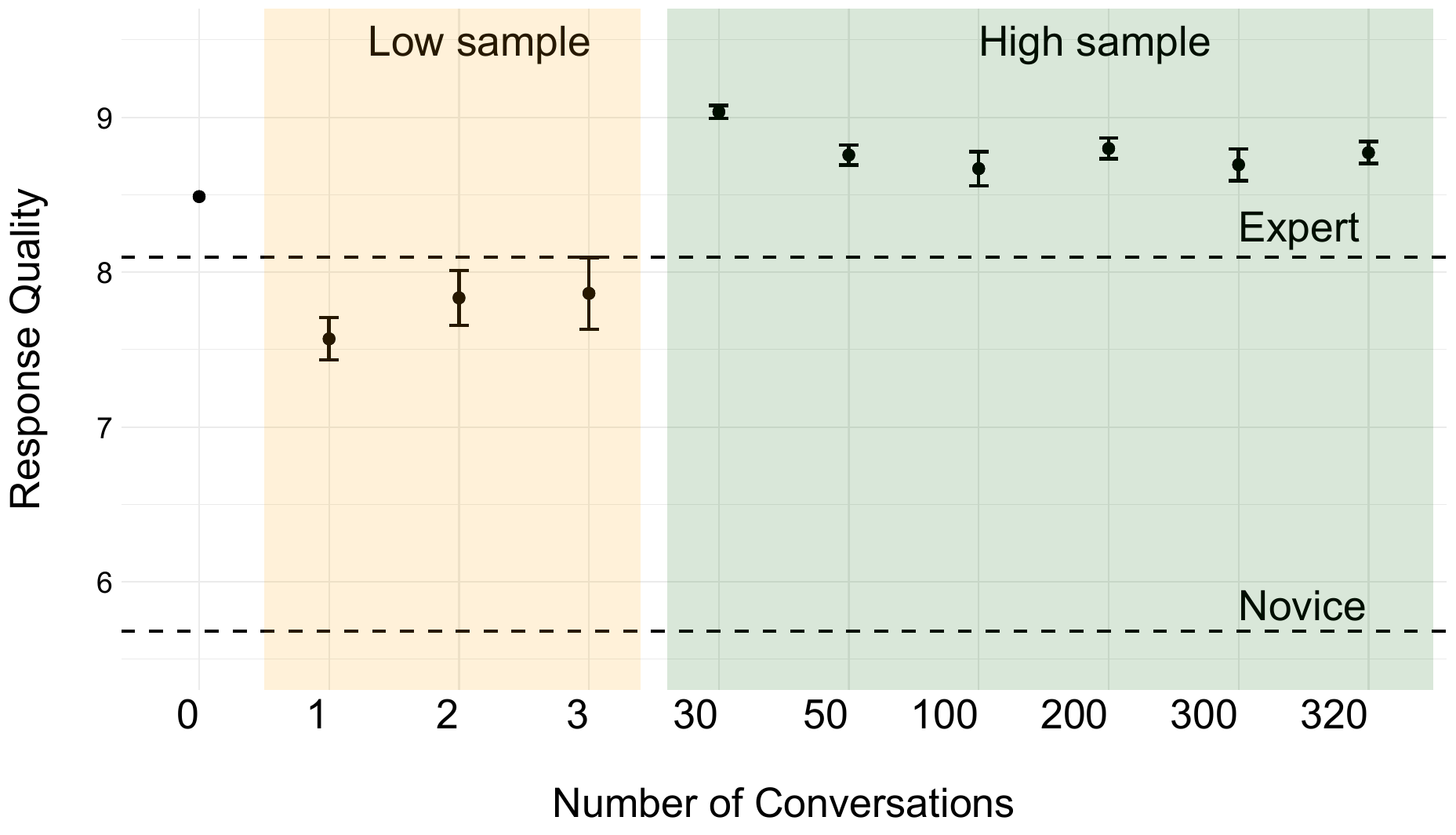}
\caption{Response quality for different numbers of conversations during knowledge externalization (95\% CIs). GPT-4o, a large language model (LLM), externalizes the tacit knowledge and generates the responses in the sample-size conditions. The horizontal axis gives the number of example tutoring conversations available during tacit knowledge externalization. At zero conversations, \textit{LLM Baseline} externalizes without example conversations; the resulting tacit knowledge is still supplied during response generation. \textit{No Tacit Knowledge}, by comparison, supplies no added guidance. At positive sample sizes,  tacit knowledge is externalized using the \textit{LLM+Conversation} prompt, which uses tutoring conversations containing novice and expert responses without a three-step requirement. Smaller samples are constructed by nested subsampling from the 320 training conversations. The dashed lines labeled ``Expert'' and ``Novice'' show scores for the original expert- and novice-written responses to the same testing conversations.}
\label{fig:robustness_subsample}
\end{figure}

Under a given sample size, we use the prompt from the {\it LLM+Conversation} condition for knowledge externalization and repeat externalization and response generation ten times using the same set of conversations. Same as before, the quality of generated responses is evaluated on the set of 122 ``complete'' testing conversations. Figure \ref{fig:robustness_subsample} reports the average response quality across different sample sizes. The expert and novice response quality scores for the same testing conversations are marked with dashed lines. The score corresponding to zero conversations is computed based on responses from the {\it LLM Baseline} condition, for which no expert conversations were provided during knowledge externalization.

Even when very few conversations are provided (between 1 and 3 conversations, low sample region highlighted in yellow), response quality already outperforms that of novices. However, response quality is relatively low and unstable, falling below or near the expert benchmark. In other words, minimal exposure to expert behavior is insufficient to extract tacit knowledge that can generate effective responses. However, with sufficient exposure to conversations (high sample region highlighted in green), response quality consistently exceeds the expert benchmark and stabilizes at a high level, and continually increasing sample sizes does not necessarily increase response quality.

These results provide evidence that LLMs are not merely mimicking expert responses but are learning decision patterns from expert conversations. The presence of a ``learning curve'' suggests that for effective knowledge externalization and response generation, the model needs exposure to a sufficient number of expert conversations. Beyond a threshold, additional conversations do not necessarily lead to further gains, indicating that the model has already captured the core of experts' tacit knowledge.

\subsection{Analysis with Alternative Tacit Knowledge}
To further validate that response quality is driven by effective tacit knowledge externalization, we replace the LLM-externalized knowledge in Study 1 with several alternative tacit knowledge representations. We consider three alternatives: (i) {\it domain-agnostic tacit knowledge}, which are tutoring tips that can be applied to any discipline, (ii) {\it out-of-context tacit knowledge} from a different discipline (namely tutoring tips on writing), and (iii) {\it domain-adapted tacit knowledge} (where writing-related key terms in (ii) are mechanically replaced with math-related vocabulary). We use each of these alternative tacit knowledge to generate responses and evaluate their quality. We construct two variants of each type of alternative tacit knowledge, and details about their sources are explained in the Materials and Methods section. The contents of each alternative tacit knowledge are reported in the {\it SI Appendix}, Table S7. 

\begin{figure*}[!t]
\centering
\includegraphics[scale=0.3]{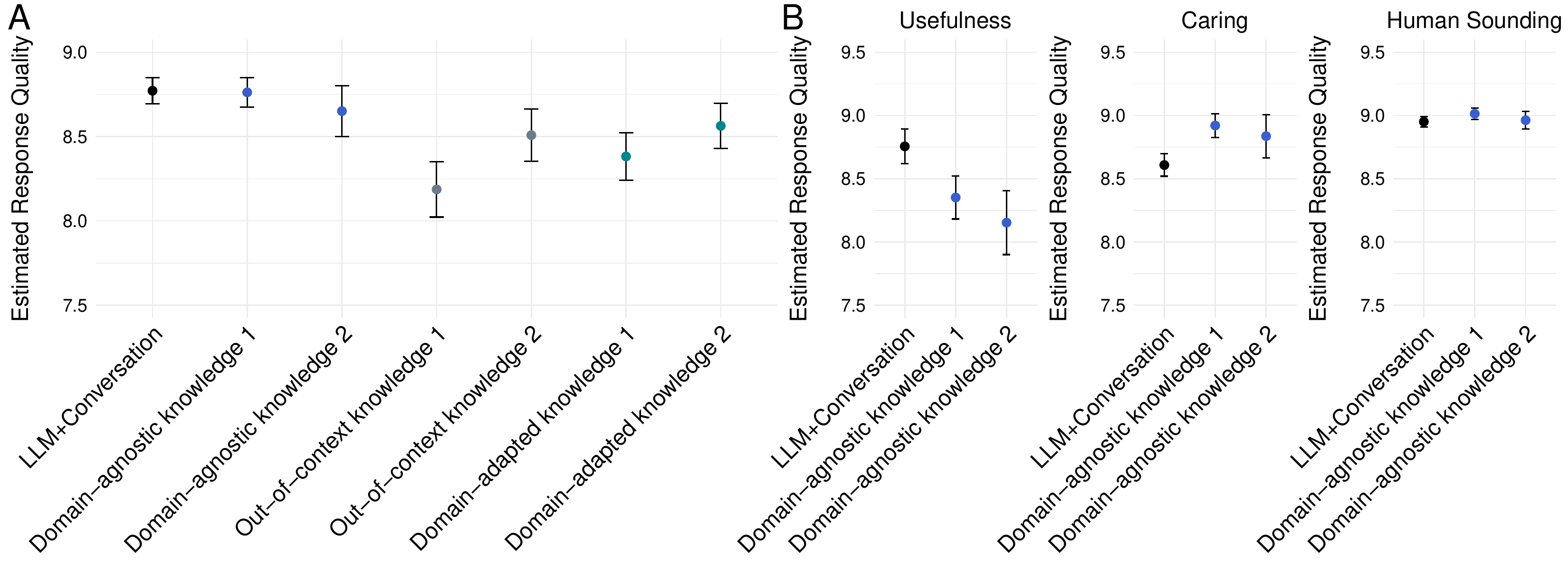}
\caption{Response quality using alternative tacit knowledge (95\% CIs). All responses were generated by GPT-4o, a large language model (LLM). Panel A reports the quality of responses generated using in-domain tacit knowledge (\textit{LLM+Conversation}) and three alternative forms of tacit knowledge. \textit{LLM+Conversation} externalizes tacit knowledge from tutoring conversations containing novice and expert responses, without a three-step requirement. The alternatives are (i) domain-agnostic tacit knowledge, comprising tutoring tips applicable to any discipline; (ii) out-of-context guidance, comprising tutoring tips from writing; and (iii) domain-adapted guidance, in which writing-related key terms are mechanically replaced with math-related vocabulary. Panel B decomposes response quality for \textit{LLM+Conversation} and the two domain-agnostic guides into usefulness, caring, and human-sounding ratings.}
\label{fig:Robustness_badmap_combined}
\end{figure*}

We repeated the response generation process from Study 1 after replacing the LLM-externalized knowledge with each of these alternative tacit knowledge representations. We used the same LMM to predict response quality from the fixed effect of condition, with conversation ID included as a random intercept. Panel A in Figure \ref{fig:Robustness_badmap_combined} reports the results. {\it LLM+Conversation} corresponds to that in Figure \ref{fig:response_generation}, representing tacit knowledge externalized by an LLM, given domain-relevant information (i.e., in-domain tacit knowledge).

Compared to all alternative tacit knowledge conditions, {\it LLM+Conversation} generates responses of the highest quality. Out-of-context tacit knowledge performed worse than domain-agnostic tacit knowledge. Replacing writing-domain terminology with math-related vocabulary, as a naive attempt at domain adaptation, marginally improves response quality but remains inferior to in-domain LLM-externalized knowledge. These results support that effective response generation depends on tacit knowledge containing domain-relevant information. Furthermore, surface-level domain adaptation in knowledge representation may not be sufficient to generate quality responses.

Tukey's HSD tests show that while {\it LLM+Conversation} significantly outperforms both variants of out-of-context tacit knowledge ({\it Out-of-context knowledge 1}: $\Delta M$ = 0.585, \(p < 0.0001\), {\it Out-of-context knowledge 2}: $\Delta M$ = 0.263, \(p < 0.01\)) and domain-adapted tacit knowledge ({\it Domain-adapted knowledge 1}: $\Delta M$ = 0.390, \(p < 0.0001\), {\it Domain-adapted knowledge 2}: $\Delta M$ = 0.208, \(p < 0.05\)), its difference with domain-agnostic tacit knowledge is not statistically significant ({\it Domain-agnostic knowledge 1}: $p = 1.00$, {\it Domain-agnostic knowledge 2}: $p = 0.608$).

While this seems to suggest that LLM-externalized domain-specific tacit knowledge does no better than domain-agnostic tacit knowledge, deeper analyses reveal an interesting nuance. We decomposed the total quality score into its three components. Panel B in Figure \ref{fig:Robustness_badmap_combined} compares the response quality for {\it LLM+Conversation} and the two domain-agnostic tacit knowledge for each component. Tukey's HSD tests show that {\it LLM+Conversation} achieves the highest usefulness scores, significantly outperforming both variants of domain-agnostic tacit knowledge ({\it Domain-agnostic knowledge 1}: $\Delta M$ =  0.404, \(p < 0.0001\), {\it Domain-agnostic knowledge 2}: $\Delta M$ = 0.603, \(p < 0.0001\)). A different pattern emerges for caring. Domain-agnostic tacit knowledge performs strongly, one even exceeding the in-domain LLM-externalized knowledge ({\it Domain-agnostic knowledge 1}: $\Delta M$ = 0.312, \(p < 0.0001\), {\it Domain-agnostic knowledge 2}: $\Delta M$ = 0.228, \(p = 0.072\)). Human-sounding scores are uniformly high across all conditions, although differences exist (e.g., {\it Domain-agnostic knowledge 1} outperforms {\it LLM+Conversation}; \(p < 0.0001\)).

These results show that the advantage of in-domain LLM-externalized tacit knowledge (over domain-agnostic tacit knowledge) is driven primarily by improvements in usefulness rather than caring or human-sounding. Caring and human-sounding appear less sensitive to domain specificity, and domain-specific tacit knowledge does not necessarily outperform domain-agnostic knowledge. Compared to domain-agnostic knowledge, the main advantage of {\it LLM+Conversation} is the improvements in instructional guidance that help students learn from their mistakes.

\vspace{-6pt}

\subsection{Replication with Retrieval-Augmented Generation (RAG)}

In the main analysis, knowledge externalization in the {\it LLM+Conversation} and {\it LLM+Both} conditions uses a subsample of 320 training conversations due to GPT-4o's context window limitations. As a robustness check, we repeat the knowledge externalization and response generation using all 420 training conversations. Instead of putting the conversations in the prompt, we embedded the training conversations into OpenAI’s vector database and had the LLM reference them via Retrieval-Augmented Generation (RAG) \cite{lewis_retrieval-augmented_2021}.

{\it SI Appendix}, Figure S12's Panel A shows the quality of responses across the six conditions. The change in knowledge retrieval method affects {\it LLM+Conversation} and {\it LLM+Both}, which now leverage the full training set of 420 conversations. Response quality for other conditions remains identical to those in Figure \ref{fig:response_generation}. {\it SI Appendix}, Figure S13 breaks down response quality into usefulness, caring, and human-sounding. The results are consistent with the main analysis.

We also repeat the analysis with alternative tacit knowledge, using the in-domain tacit knowledge externalized using the full set of 420 training conversations. Results are shown in {\it SI Appendix}, Figure S14. In-domain knowledge (i.e., {\it LLM+Conversation}) generates responses of the highest quality, and Tukey's HSD comparisons show that it significantly outperforms all alternative tacit knowledge conditions ($p < 0.01$). We then decompose the response quality into its components. Consistent with the main analysis, Tukey's HSD comparisons show that {\it LLM+Conversation} significantly outperforms both domain-agnostic tacit knowledge in terms of usefulness ($p < 0.001$). {\it LLM+Conversation} performs similarly to domain-agnostic tacit knowledge in terms of caring ({\it Domain-agnostic knowledge 1}; $p = 0.069$, {\it Domain-agnostic knowledge 2}; $p = 0.217$) and human-sounding ({\it Domain-agnostic knowledge 1}; $p =
0.081$, {\it Domain-agnostic knowledge 2}; $p = 0.994$). 

\vspace{-6pt}

\subsection{Replication with Other LLMs}

For our main analysis, we used GPT-4o throughout the entire experiment pipeline: knowledge externalization, response generation, and response evaluation. To check the robustness of our results, we replicate our analysis using two different models. First, we repeat the response evaluation step for Study 1 and Study 2 using xAI's Grok model. Then, we repeat the entire pipeline using Anthropic's Claude models. The models used in this section are summarized in {\it SI Appendix}, Table S8. Table \ref{table_Claude} reports consistency across GPT-4o and other models on key results.

First, we use xAI's Grok 4 Fast (non-reasoning) to evaluate response quality. The externalized knowledge and generated responses are the ones created by GPT-4o in the main analyses. {\it SI Appendix}, Figure S15 shows that for Study 1, results are generally consistent with those evaluated using GPT-4o. Main findings hold when using Grok as the scoring model. We break down the response quality scores for Grok evaluation into the three components. {\it SI Appendix}, Figure S16 is qualitatively similar to {\it SI Appendix}, Figure S10. 

For Study 2, using Grok as the evaluating LLM results in similar patterns as those using GPT-4o. {\it SI Appendix}, Figure S17 shows consistent patterns as Figure \ref{fig:knowledge_transfer_inverted}. Breaking down the response quality scores, we see a similar pattern between {\it SI Appendix}, Figure S18 and {\it SI Appendix}, Figure S11. 

To further assess the robustness and generalizability of our results, we replicate the entire analysis using LLMs from a different provider. We use Claude Haiku 4.5 (released October 2025, knowledge cutoff February 2025) and Sonnet 4.5 (released September 2025, knowledge cutoff January 2025). For each model, we implement the complete pipeline of knowledge externalization, response generation, and response evaluation.

{\it SI Appendix}, Figure S19's Panel A and B results are largely consistent with the GPT-based analysis. Across models, the \textit{LLM+Conversation} condition consistently performs the best or one of the best. This reinforces the importance of providing expert behavior in externalizing tacit knowledge. Table \ref{table_Claude} reports consistency between the main findings from GPT-4o and Claude-based analyses.

To study boundary conditions related to model capability, we conduct an exploratory analysis using a notably weaker model, Claude 3 Haiku (released March 2024, knowledge cutoff August 2023). {\it SI Appendix}, Figure S19's Panel C shows a different pattern compared to the GPT-based analysis. {\it Expert} and {\it Human Best Effort} generate higher quality responses than all LLM-externalized tacit knowledge conditions. Unlike the stronger models, tacit knowledge does not improve response quality relative to {\it No Tacit Knowledge}. Tukey's HSD tests show that there is no statistically significant difference between {\it No Tacit Knowledge} and {\it LLM Baseline} ($p = 0.973$) and {\it LLM+Conversation} ($p = 0.931$), and {\it LLM+Both} performs significantly worse ($p < 0.001$). Although {\it LLM+Conversation} remains highest among LLM-externalized knowledge conditions, it underperforms {\it Expert} ($p < 0.001$) and {\it Human Best Effort}  ($p < 0.01$).

These results suggest that the effectiveness of tacit knowledge externalization depends on model capability. A sufficiently strong LLM is needed to be able to extract and apply tacit knowledge from expert behavior, whereas a weak LLM would fail to realize these benefits.

\begin{table}[t!]
\centering
\caption{Robustness of main findings of Study 1 across models}
\label{table_Claude}
\renewcommand{\arraystretch}{1}
\small

\begin{tabular}{p{0.30\columnwidth} 
                p{0.07\columnwidth} 
                p{0.07\columnwidth} 
                p{0.14\columnwidth} 
                p{0.14\columnwidth}}
\hline
\\[-1.8ex]
\multicolumn{1}{c}{\textbf{Finding}} & 
\multicolumn{1}{c}{\textbf{GPT}} &
\multicolumn{1}{c}{\textbf{Grok}} &
\multicolumn{1}{c}{\textbf{Haiku 4.5}} & 
\multicolumn{1}{c}{\textbf{Sonnet 4.5}}  
 \\
\hline
\\[-1.8ex]

Tacit knowledge improves response quality relative to \textit{No Tacit Knowledge}  &

\multicolumn{1}{c}{Yes}  & 

\multicolumn{1}{c}{Yes}  & 

\multicolumn{1}{c}{Yes}  & 

\multicolumn{1}{c}{Yes}  

\\[1ex]

LLM-externalized tacit knowledge conditions outperform \textit{Expert}  &

\multicolumn{1}{c}{Yes$^{\S}$}  &

\multicolumn{1}{c}{Yes$^{\S}$}  &

\multicolumn{1}{c}{Yes}  & 

\multicolumn{1}{c}{Yes}  

\\[1ex]

\textit{LLM+Conversation} outperforms \textit{Human Best Effort}  &

\multicolumn{1}{c}{Yes}  &

\multicolumn{1}{c}{Yes}  &

\multicolumn{1}{c}{Yes}  & 

\multicolumn{1}{c}{Yes}

\\[1ex]

\textit{LLM+Conversation} produces the highest-quality responses &

\multicolumn{1}{c}{Yes} & 

\multicolumn{1}{c}{Yes} & 

\multicolumn{1}{c}{Yes$^{+}$} & 

\multicolumn{1}{c}{Yes$^{+}$}

\\[1ex]

\hline

\vspace{-0.4em}

\parbox{0.7\columnwidth}{
\footnotesize {

$^{\S}$ {\it LLM+Conversation} and {\it LLM+Both} significantly outperform {\it Expert}, and there is no statistically significant difference between {\it Expert} and {\it LLM Baseline}

$^{+}$ No significant difference between the three LLM-externalized conditions.
}
\vspace{0.5em}

This table summarizes whether results using alternative large language models (LLMs) replicate directionally and statistically relative to the GPT-4o baseline. The third column reports results using Grok 4 Fast (non-reasoning) as the evaluating model instead of GPT-4o. The fourth and fifth columns report results in which Claude Haiku 4.5 and Sonnet 4.5 replicate the entire pipeline of knowledge externalization, response generation, and response evaluation. Condition names identify the tacit knowledge supplied: \textit{No Tacit Knowledge}, no added guidance; \textit{Human Best Effort}, the three-step framework developed through interviews with expert tutors in prior research; \textit{LLM Baseline}, externalized by the LLM without example tutoring conversations; \textit{LLM+Conversation}, externalized from tutoring conversations containing novice and expert responses, without a three-step requirement; and \textit{LLM+Both}, externalized from such conversations with a three-step requirement. \textit{Expert} denotes the original expert-written responses.}

\end{tabular}
\end{table}

\subsection{Validation of Evaluation Criteria Using Experienced Human Tutors}

To assess whether LLMs can serve as reliable evaluators of response quality, we conducted an evaluation study comparing LLM-generated scores with scores provided by human evaluators with prior math tutoring experience. We recruited eight experts with math tutoring experience (four undergraduate students and four PhD students) with an average of 35 months of math tutoring experience. Each undergraduate tutor evaluated 20 responses, and each PhD tutor evaluated 40 responses on each of the three criteria. Further details of the study are described in the Materials and Methods section. Because our analyses primarily focus on relative comparisons across responses, we evaluate consistency using Spearman rank correlations. We additionally report Pearson correlations to assess linear association in absolute scores.

Spearman correlation scores show strong agreement between LLM and human evaluations for overall response quality ($\rho = 0.707$, $p < 0.001$), indicating substantial consistency in the relative ordering of responses. Examining the three evaluation dimensions separately, usefulness shows moderate agreement ($\rho = 0.541$, $p < 0.001$), caring shows strong agreement ($\rho = 0.787$, $p < 0.001$), and human-sounding shows moderate agreement ($\rho = 0.533$, $p < 0.001$) \cite{dancey_statistics_2007, akoglu_users_2018}.
Pearson correlations show a similar pattern of results, with correlation coefficients for overall response quality ($r = 0.775$, $p < 0.001$), usefulness ($r = 0.618$, $p < 0.001$), caring ($r = 0.866$, $p < 0.001$), and human-sounding ($r = 0.598$, $p < 0.001$). Together, these results suggest that LLM-based evaluations are generally well aligned with experienced tutors' judgments.

\vspace{-6pt}

\section{Discussion}

Tacit knowledge externalization and transfer remain a persistent challenge in organizations because experts often cannot accurately articulate the knowledge underlying their behaviors and decisions \cite{agarwal_knowledge_1990, collins_chapter_2008}. As a result, tacit knowledge is difficult to document, limiting its dissemination and even risking oblivion when experts leave the organization \cite{kransdorf_corporate_1998, howells_tacit_1996}. This study investigates whether LLMs can address this challenge by externalizing tacit knowledge based on experts' behaviors, and whether such externalized knowledge supports downstream decision-making and knowledge transfer. In a math tutoring task, we show that LLMs can effectively extract tacit knowledge from expert behavior, and this improves decision quality and helps reduce the expert-novice score gap.  

In Study 1, LLM-externalized tacit knowledge derived from expert conversations substantially improves downstream response quality, beyond expert-articulated tacit knowledge. We also find that while providing expert conversations improves response quality, imposing structural restrictions to mimic expert-articulated knowledge reduces it. This pattern suggests that allowing output flexibility is important for LLMs to capture the nuances of expert tacit knowledge embedded in observed behaviors.  

Study 2 demonstrates that LLM-externalized tacit knowledge can facilitate knowledge transfer to novices. Novices trained on LLM-externalized tacit knowledge based on expert conversations show a significantly reduced expert–novice gap. Notably, tacit knowledge based on expert conversations outperforms expert-articulated tacit knowledge. The improvement in novices' response quality within this brief experiment highlights the practical potential of using LLMs for tacit knowledge transfer.

Our mechanism analyses suggest that LLMs externalize tacit knowledge that improves different dimensions of response quality. Relative to {\it Human Best Effort}, {\it LLM+Conversation} produced more caring and human-sounding responses in Study 1 and more caring novice responses in Study 2. The usefulness differences were not statistically significant in either study. These findings suggest that the clearest advantages of conversation-derived tacit knowledge concern how responses are communicated 

Our robustness analyses show that LLMs meaningfully {\it learn} and {\it extract} tacit knowledge from expert conversations. First, we see a ``learning curve'', where learning from very few examples does not generate high-quality responses, but once sufficient examples are provided, LLM-externalized tacit knowledge leads to better responses than expert-articulated tacit knowledge. This suggests that LLMs are not merely mimicking expert responses but are meaningfully learning decision patterns underlying them.

Additionally, in-domain LLM-externalized tacit knowledge generates higher-quality responses than alternative representations of tutoring knowledge that are either out-of-context or domain-adapted. Compared to domain-agnostic tacit knowledge, the advantage of in-domain LLM-externalized tacit knowledge lies primarily in usefulness rather than caring. This suggests that the value of LLM-externalized tacit knowledge lies in its ability to externalize instrumental domain-specific knowledge.

These findings are robust across LLMs and knowledge retrieval methods used for knowledge externalization (in-context learning or vector search). However, a weak LLM (e.g., Claude 3 Haiku) is less effective at externalizing tacit knowledge from expert behaviors and applying it to response generation, highlighting an important boundary condition.

This work contributes to the literature on knowledge management and organizational learning by introducing a scalable approach to tacit knowledge externalization. Prior approaches have relied on knowledge elicitation methods \cite{clark_cognitive_2008,ryder_integrating_1993,wang_bridging_2024, lebovitz_is_2021, bradley_analyzing_2006, cooke_varieties_1994}, relying on experts' limited ability to verbalize their own reasoning. Moreover, these methods are resource-intensive and difficult to scale. In contrast, we show that LLMs can infer tacit knowledge from experts' behavior and also demonstrate that this method effectively supports downstream decision-making and knowledge transfer. These findings build upon prior research on LLM-enabled knowledge externalization (e.g., \cite{zuin_leveraging_2025,kuks_using_2025, wang_why_2025, lu_tacit_2025, schinckus_large_2026}) by demonstrating not only that LLMs can uncover patterns from expert behavior, but also that such externalized knowledge has downstream utility. 

Our findings also empirically reaffirm Polanyi’s Paradox \cite{polanyi_tacit_1966} while also suggesting a way to mitigate it. LLM-externalized tacit knowledge based on expert conversations consistently outperforms expert-articulated tacit knowledge in both response generation and knowledge transfer. These findings suggest that although experts possess rich tacit knowledge, they may struggle to fully articulate it. As a result, expert-articulated tacit knowledge may fail to fully capture the tacit knowledge that experts use. LLMs can mitigate this articulation bottleneck by directly learning from experts' observed behaviors that contain their tacit knowledge. 

This study offers practical implications for organizational training and decision support. Organizations possess large repositories of expert behavioral data, such as customer support chats, client meeting logs, etc. \cite{ziora_role_2015}. Our findings suggest that organizations can leverage such data to externalize experts' tacit knowledge. For example, organizations can use LLM-externalized tacit knowledge to prompt LLMs to provide decision support for employees (e.g., recommending actions) and to facilitate novice training. Importantly, our findings suggest that effective externalization depends on providing the LLM with experts' observed behavior while allowing output flexibility. It also depends on providing the LLM with sufficient exposure to expert behavior, as well as using sufficiently capable models.

While this study provides important insights, several limitations offer directions for future research. First, our empirical setting is limited to a single context --- math tutoring --- and a specific form of expert behavior data -- tutoring conversations. Math tutoring is a domain in which expert behavior is relatively observable and structured. Future studies could examine whether these findings generalize to other domains in which expert behavior is less structured or observable, such as sales, negotiation, and creative professions. Second, our knowledge transfer experiment captures relatively short-term learning effects. Although novices exposed to LLM-externalized tacit knowledge showed meaningful improvements even within a brief intervention period, future research should investigate whether such gains persist over the long term.

\bibliographystyle{apalike}
\bibliography{arxiv_ref}

\clearpage
\includepdf[pages=-]{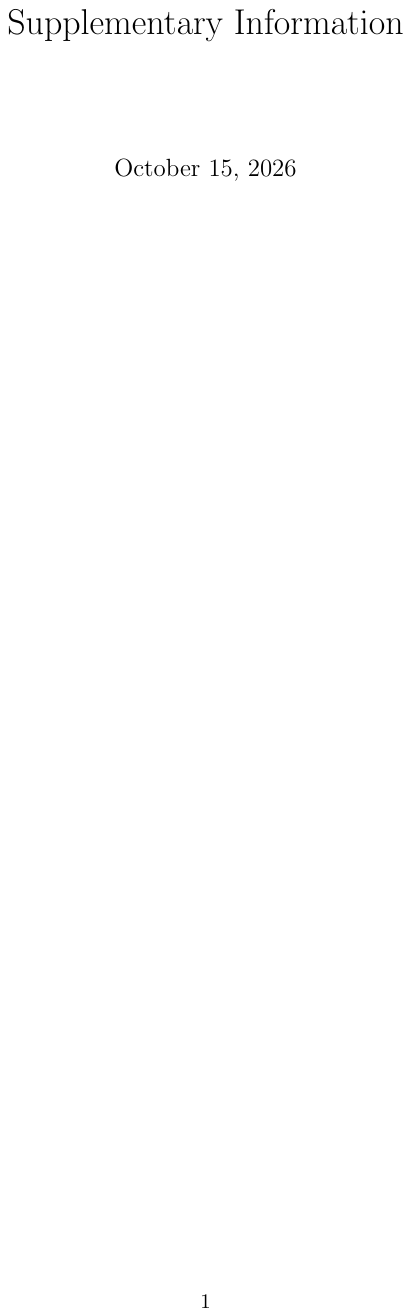}

\end{document}